\documentclass[preprint]{aastex63}

\usepackage{xcolor}
\usepackage{enumerate}

\let\a=\alpha \let\b=\beta \let\g=\gamma \let\d=\delta \let\e=\epsilon
\let\h=\eta  \let\k=\kappa \let\l=\lambda 
\let\s=\sigma \let\t=\tau \let\u=\upsilon
\let\r=\rho
 
\let\w=\omega	   \let\G=\Gamma \let\D=\Delta  \let\L=\Lambda
\let\S=\Sigma \let\W=\Omega
\let \x = \chi \let \z = \xi

\def\size{\displaystyle}

\def\be{\begin{eqnarray}}
\def\ee{\end{eqnarray}}
\def\ba{\begin{array}}
\def\ea{\end{array}}

\def \yr {\mathrm{yr}}
\def \Jy {\mathrm{Jy}}
\def \arcsec {\mathrm{arcsec}}

\def \kms {\mathrm{km \, s^{-1}}}
\def \cm {\mathrm{cm}}
\def \M  {\mathscr{M}}
\def \K {\mathrm{K}}
\def \sec {\mathrm{s}}

\def\sch{\mathrm{s}}

\usepackage{mathrsfs}  
\def \M  {\mathscr{M}}

\shorttitle{Neutral gas in the accretion zone of Sgr A*}
\shortauthors{Murchikova, Wang, Mason, Blandford}

\graphicspath{{./}{figures/}}

\begin{document}

\title{Neutral Gas within 20,000 Schwarzschild radii of Sagittarius A*}

\correspondingauthor{Elena (Lena) Murchikova}
\email{lena@ias.edu}

\author[0000-0001-8986-5403]{Elena M. Murchikova}
\affiliation{Institute for Advanced Study, 1 Einstein Drive, Princeton, NJ 08540, USA}

\author[0000-0002-0042-9873]{Tianshu Wang}
\affiliation{Department of Astrophysical Sciences, Princeton University, Princeton, NJ 08544, USA}

\author[0000-0002-8472-836X]{Brian Mason}
\affiliation{National Radio Astronomy Observatory 520 Edgemont Road Charlottesville, VA 22903, USA}

\author[0000-0002-1854-5506]{Roger D. Blandford}
\affiliation{Kavli Institute for Particle Astrophysics and Cosmology, Stanford University, Stanford, CA 94309, USA}

\begin{abstract}

\citet{Murchikova2019} 
discovered a disk of cool ionized gas within 20,000 Schwarzschild radii of the Milky Way's Galactic Center black hole Sagittarius A*. They further demonstrated that
the ionizing photon flux in the region is enough to keep the disk ionized, but there is not ample excess of this radiation. This raised the possibility that some neutral gas could also be in the region shielded within the cool ionized clumps. Here we present ALMA observations of a broad 1.3 millimeter hydrogen recombination line H30$\alpha$: $\mathrm{n} = 31 \to 30,$ conducted during the flyby of the S0-2 star by Sgr A*. We report that the velocity-integrated H30$\a$ line flux two month prior to the S0-2 pericenter passage is about 20\% larger than it was one month prior to the passage. The S0-2 is a strong source of ionizing radiation moving at several thousand kilometers per second during the approach. Such a source is capable of ionising parcels of neural gas along its trajectory, resulting in variation of the recombination line spectra from epoch to epoch.
We conclude that there are at least $(6.6 \pm 3.3) \times 10^{-6} M_\sun$ of neutral gas within 20,000 Schwarzschild radii of Sgr~A*.

\end{abstract}

\keywords{
Supermassive black holes (1663); Low-luminosity active galactic nuclei (2033); Galactic center (565)
}

\section{Introduction}

Galactic Center black hole  Sagittarius A* (Sgr A*) is surrounded by a hot X-ray emitting accretion flow, which is radiatively inefficient 
\citep{Narayan1994,Blandford1999,Quataert2000,Tchekhovskoy2012, Ressler2018}. 
Observational constraints on the properties of this flow at large scales ($10^5$ Schwarzschild radii ($R_\sch$) $\simeq$ 0.05 pc from the black hole and greater) primarily come from observations of hot gas (around $10^7$ K) \citep{Baganoff2003}. At near horizon scales, we rely on polarization measurements \citep{Bower2003} and soon the Event Horizon Telescope observations. At intermediate scales, there are too few, if any, model independent probes to reliably constrain physical properties. For a long time, the task of interpolating between large and small scales has been left to numerical simulations, which tend to focus on X-ray emitting gas, while ignoring the cool phase (defined here as the gas at temperatures about $10^4$ K gas and cooler).

In \citealt{Murchikova2019}, we discovered a disk of cool gas at an intermediate distance from the black hole (within 0.23 arcsec or $0.009$~pc radius) in the 1.3 mm recombination line of hydrogen H30$\alpha$ observed with the Atacama Large Millimeter/submillimeter Array (ALMA). This demonstrated that cool gas can survive quite close to the black hole. The disk with the hole in the middle is composed of a collection of cloudlets.
The detection of a double peaked H30$\alpha$ line was later confirmed by \cite{Yusef-Zadeh2020}.
Simulations of \cite{Calderon2020} reported the formation of a similar cool disk from colliding of stellar winds.

The existence of relatively cool and dense ionized gas clumps within $\sim 2 \times 10^4 R_\sch$ of the black hole raises the possibility that the neutral component could also be present, shielded inside the ionized gas clumps. The contribution of the neutral component of the accretion flow at $\sim 10^4 \, R_\sch$ is so far unconstrained, as the neutral gas is invisible in both X-rays and recombination lines. 
 
In order to constrain the amount of the neutral gas in the region we need to see it. If it could be ionized by a fast-moving object for a relatively short period of time it would become visible in recombination lines. In the spring of 2018, this is exactly what happened: the S0-2 (S2) star made its closest approach to Sgr A*, providing a fast-moving strong source of ionizing radiation. During the approach it moved with a velocity of several thousand $\kms$ and came as close as 1200 $R_\sch$ \citep{GRAVITY2018,Do2019} to the black hole. S0-2  is an early B star of spectral type B0-B2.5V which produces about $4.7\times 10^{46}$ photons, having energies greater than 13.6 eV per second (Appendix \ref{app:S0-2}). If pockets of neutral gas existed within the cool disk, S0-2 would have ionized those in its vicinity. After the star's departure, the neutral gas in the pockets would have started to recombine, producing an excess of recombination line emission. As a result, the spectrum of the cool disk observed in H30$\alpha$ recombination line would be expected to change with time as a function of the location of the S0-2 star.
The absence of recombination line excess would indicate the absence of the neutral gas in the cloudlets of the cool disk.

Here we report ALMA observations of the hydrogen recombination line H30$\alpha: \mathrm{n}=31 \to 30$  from the cool disk \citep{Murchikova2019} during the close approach of the S0-2 star. Our data were obtained two months and one month before the star's pericenter passage -- in March and in April 2018, respectively. We find an increase of the velocity-integrated line flux between the 2016 observations (M19) and the 2018 observations presented here, and substantial variation between the March and April 2018 observations. Namely, in June-August 2016 the velocity-integrated line flux was $3.8 \, \Jy \, \kms,$ around March 15, 2018 it increased to $6.1 \, \Jy \kms,$ and around April 15, 2018 it decreased to $4.9 \, \Jy \, \kms.$  

The change in the velocity-integrated line flux between 2016 and 2018 could be attributed to various causes: (i) the rotation of a geometrically asymmetric distribution of cloudlets, (ii) the disk's slow evolution due to evaporation of cool and condensation of hot has phases, and (iii) winds of the star blowing away parts of the disk which were directly in front of the star. We will discuss the global evolution of the cool disk spectrum (using observations conducted in 2016, 2017, and 2018) in Murchikova et al 2022 (in preparation).

In this work we focus on the month-scale variability. We interpret the change of the velocity-integrated line flux between mid-March and mid-April 2018 as a signature of neutral hydrogen gas parcels present in the cool disk. Recombination time of the neutral gas in such parcels would be about one month, which is consistent with the timescale of the observed variability.

This paper is organized as follows. Section \ref{sec:obs} describes our observations. Section \ref{sec:analysis} presents the details of the data analysis and spectral extraction. We discuss the results in Section \ref{sec:results} and their physical implications in Section \ref{sec:discussion}. We conclude in Section \ref{sec:conclusion}. Appendix \ref{app:S0-2} is dedicated to derivation of the ionizing photon flux from the S0-2 star. Appendix \ref{app:dip} describes the subtraction of spectral contamination by emission from large radii. Appendix \ref{app:coord} presents the rotational matrices of coordinate transformations from the orbital plane of the S0-2 star to the RA and Dec on the sky.

\section{Observations and Data Reduction}\label{sec:obs}

Our data were obtained during ALMA Cycle 5 for project 2017.1.00995.S (PI Murchikova) and consist of two sets: (i) {\it the March data} -- the observations conducted on March 13 and 15, 2018, which is two months before the S0-2's pericenter passage on May 19, 2018, and (ii) {\it the April data} -- the observations conducted on April 16 and 18, 2018, which is one month before the S0-2's pericenter passage.  

 
The observations were centered on Sgr A$^*$: RA 17:45:40.0359, Dec -29:00:28.169 (J2000) and conducted in receiver Band 6 in the spectral scan mode. The spectral frequency range was between 223.500~GHz and 240.000 GHz. The spectral scan had five tunings with two 1.875GHz-wide spectral windows each.  Spectral windows were centered on 
224.44 GHz, 226.14 GHz, 227.84 GHz, 229.55 GHz, 231.25~GHz, 232.95 GHz, 234.66 GHz, 236.36 GHz, 238.06 GHz, 239.77 GHz, 
and had 128 channels each. 
The correlator was configured in the time division mode (TDM).


Since the observations were tied to the orbital position of S0-2 and were therefore time-sensitive,  we used the telescope configurations available.
The March data were obtained in ALMA configuration C43-4 with 45 12-meter antennas. The beam size was $0.6 \times 0.5 \, \arcsec^2$. The integration time on target was 152 
minutes. The total observing time including calibrations was 
6 hours. The achieved sensitivity was 0.3 mJy/beam in a 60 $\kms$ velocity channel.
The April data were obtained in ALMA configuration C43-3 with 47 12-meter antennas. The beam size was $1.0 \times 0.8 \, \arcsec^2$. The integration time on target was 190 
minutes. The total observing time including calibrations was 
7.5 hours. The achieved sensitivity was 0.3 mJy/beam in a 60~$\kms$ velocity channel.

For calibration and data reduction we used the scripts manually prepared by the staff at North American ALMA Science Center (NAASC) in Charlottesville, Virginia and delivered with the data.


\section{Data Analysis}\label{sec:analysis}


Our goal is to measure the (faint and broad) H30$\alpha$ line
in the presence of a (bright and compact) 
continuum source (Sgr A*). The procedure we developed to do so is described below; all processing was carried out in CASA (version $5.6.1$) and was performed separately for each epoch (March and April).

\begin{enumerate}
\item We measured the continuum flux density of Sgr A* by fitting the visibilities in each spectral window (SPW) to a point-source (task \verb|uvmodelfit|), then used the resulting model to perform phase-only self-calibration. This procedure was iterated three times with a shorter solution interval each time; the final self-calibration solutions were derived per integration. No amplitude self-calibration was performed. The integrated flux density of the cool disk surrounding Sgr A* within 0.23 arcsec is only about $3$ mJy, or 0.1\% of the $3.5 {\rm Jy}$ flux density of Sgr A* — a negligible effect in the self-calibration. The mean flux density of the Galactic Center minispiral within 16 arcsec radius field of view is 1.5 mJy. The diameter of the black hole shadow of Sgr A* is about $5 R_\sch$ (or $3\times 10^{-5}$ arcsec), which is much less than the $0.5-1$ arcsec beam size of our observations. We can therefore safely use a point-source as a model of the continuum source.


\item \label{itm:2} In order to remove Sgr A* as a function of frequency, the \verb|uvmodelfit| was repeated for each spectral channel and the resulting model subtracted (task \verb|uvsub|). This was performed on a per execution basis.


\item \label{itm:3} We imaged the point-source-subtracted data for each epoch using \verb|tclean| with $0''.05$ pixel size. Cleaning was conducted using auto-masking (\verb|auto-multithresh2|), supplemented with interactive masking where necessary.


\item \label{itm:5} From the source-subtracted data cube for each SPW-epoch, we extracted the spectrum within a $1.2\times 0.8 \, \arcsec^2$ region centered on the position of Sgr A*. This region was slightly larger than the Half Power Beam Width (HPBW) of the lowest resolution observation we have (from mid-April). We obtained this spectrum by summing the \verb|model| points within the chosen aperture from the output of the CASA \verb|tclean| task, and adding the flux density within same aperture of the \verb|residual| image.


\item We note that the procedure in step \ref{itm:2} is an imperfect continuum subtraction,  since it will also partially subtract flux from the disk due to the large beam size of the observations. To correct for this, we took the Sgr A* spectrum from each execution (step~\ref{itm:2}, above) and estimated the continuum by fitting a linear function to it, excluding the velocity range of the expected H30$\alpha$ line (about $\pm 1000 \kms$ from 231.9 GHz). The remaining point-source spectrum was added to the diffuse source spectrum from step~\ref{itm:5} for each SPW-epoch. About 70\% of the disk flux was recovered in this fashion for the March and April data.



\item \label{itm:7} We stitched the spectra from adjacent SPWs together. ALMA spectral windows tend to be misaligned by about 0.5\% of the value of the absolute flux. The residual misalignment is in both spectral inclination and the absolute offset (detail discussion is in M19). To stitch the sides of the spectra belonging to different spectral windows, we varied the inclinations and the relative offsets. The combined spectrum is the result we analyze in this paper. 


\end{enumerate}

CASA version 5.6.1 can fit a u-v model to the data in a specified channel (task \verb|uvmodelfit|). However, it cannot write the fit to the model-column of a specified channel (task \verb|ft|), which is necessary for the fit subtraction (task \verb|uvsub|). 
To carry out this subtraction, we split the data channel-by-channel. We then fit the u-v model, subtracted it, and merged the point-source-less data from all channels to a single file.

Due to the relatively large beam size of the April observations, the u-v model fitting and subtraction of step \ref{itm:2} for this dataset is susceptible to contamination. The most obvious potential contaminant is the minispiral \citep{Tsuboi2017} and the clouds associated with it. The arms of the minispiral surround the Galactic Center black hole within 2 pc radius (50 arcsec) and its inner part approaches within one arcsecond of the black hole. The minispiral's emission is centered at the frequency of the H30$\a$ line, and has a full width of about $400 \, \kms.$ This is about five times as narrow as the cool disk emission and affects the spectra in the vicinity of the H30$\a$.

In the presence of extended emission, the task \verb|uvmodelfit| can misidentify the base level of the u-v model fitting by about 1 mJy. However this does not pose a problem. The final product of our analysis is the combined spectrum of the point-source and the extended emission (obtained from the point-source-less image). Any emission undersubtracted in step \ref{itm:2} remains in the data and is accounted for in step \ref{itm:5}. The details on removing contamination are presented in Appendix \ref{app:dip}.

The data analysis and spectrum extraction performed in this work are designed to use as little human input as possible. This is in contrast with M19, where nearly every step of the data analysis and spectrum extraction was manual.
Here we work with different observational executions separately, perform self-calibration, and remove the unresolved continuum source at the location of Sgr A* using automated u-v model fitting and subtraction. In M19 we worked with all of the observational executions combined, performed no self-calibration, and removed the continuum central source using CASA task \verb|uvcontsub| with human defined ``continuum'' channels. In a companion study (Murchikova et al 2022, in preparation) we compare these two approaches and show that they yield the same results.

In \cite{Murchikova2021b}, we quantified the variability properties of the Sgr A* point-source, and demonstrated that it is possible to extract accurate fluxes for faint point-sources in ALMA data using u-v plane modeling. That allowed us to measure Sgr A* photometry from short exposures and thereby extend the measurement of variability properties to timescales two orders of magnitude shorter than previous work. The innovative light curve extraction technique implemented in \cite{Murchikova2021b} is a derivative of the analysis developed for this work. In that paper, the authors collapse the dataset in frequency and extract the flux density variability with time using u-v model fitting  to obtain the light curves. Here, we instead collapse the dataset in time and extract the flux density variability with frequency using u-v model fitting to obtain the spectrum.

The observations were set up in the spectral scan mode and cover a range of about $\pm 10,000 \, \kms$ around the frequency of the H30$\a$ recombination line. The frequency range is covered with ten spectral windows. There were twenty overlapping channels between successive spectral windows -- ten on each end. However, due to the edge effect of the spectrometers in Time Division Mode the 9-10 edge channels are unusable. This leaves us no overlapping spectral channels to perform independent alignment of spectral windows. In-depth discussion of the misalignment of ALMA's spectral windows is presented in the Supplementary Information of \cite{Murchikova2019}.

Because of the absence of overlapping channels in the spectral scan mode and the persistence of the issue of the misaligned spectral windows, we can only align the data around the frequencies where we can use M19 spectra as reference, i.e. spectral windows 4 and 5 centered on 229.55 GHz and 231.25 GHz. For this frequency range we establish that the data are indeed consistent with the presence of a wide line around the frequency of H30$\a.$ The full width of the cool disk spectrum presented here is about the same as in M19. We see no indication that the H30$\a$ line extends past the observed $\pm 1000 - 1500 \, \kms$ (Figure \ref{fig:spec}).

The spatial diameter of the cool disk as determined in M19 is 0.45 arcsec. The beam size of the March observations is 0.5 arcsec which is about the size of the disk. The beam size of the April observations is 1.0 arcsec which is about twice the size of the disk. No imaging which would highlight structural information about the disk is possible for the dataset used in this study. 
Due to the large size of the beam compared to the size of the cool disk, about 70\% the cool disk emission is subtracted with the point-source spectrum in step \ref{itm:3}. The rest of the emission comes from the spectral extraction from the Sgr A*-less image (step \ref{itm:5}).

The dominant uncertainties of the velocity-integrated line fluxes of the March and April 2018 data come from the alignment of spectral windows. We estimate the uncertainty due to the relative inclination (step \ref{itm:7}) of the spectral windows at 10\%, and due to absolute vertical offset at  10\%. Observational uncertainty contributes $\sim 3 \%.$  
The uncertainty due to possible unaccounted for leakage emission from the region outside the cool disk (Appendix \ref{app:dip}) is about 5\%.
The uncertainty due to possible combination of narrow absorption and emission foreground molecular features is about 5\% (M19). 
We evaluate the combated uncertainties of the velocity-integrated line fluxes for the presented data at about 18\%.

One of the most striking features of the data presented here is that the mid-March 2018 spectrum of the H30$\a$ emission is stronger than it is in mid-April 2018. This feature is real. The average continuum flux during mid-March observations was 3.3 Jy and during mid-April observations was 3.6 Jy. This rules out the possibility of gain calibration error, which would imply that the disk flux scales with the continuum flux. Additionally, we made a virtual face-to-face visit to the North American ALMA Science Center at NRAO in Charlottesville to double-check the quality of the reduction and calibration. No anomalies were identified, and we concluded that the difference in the flux density between the observations is indeed physical.


\section{Results} \label{sec:results}

\begin{figure}
\vspace{-0.0cm}
\centering
\begin{tabular}{cc}
\includegraphics[width=0.7\columnwidth]{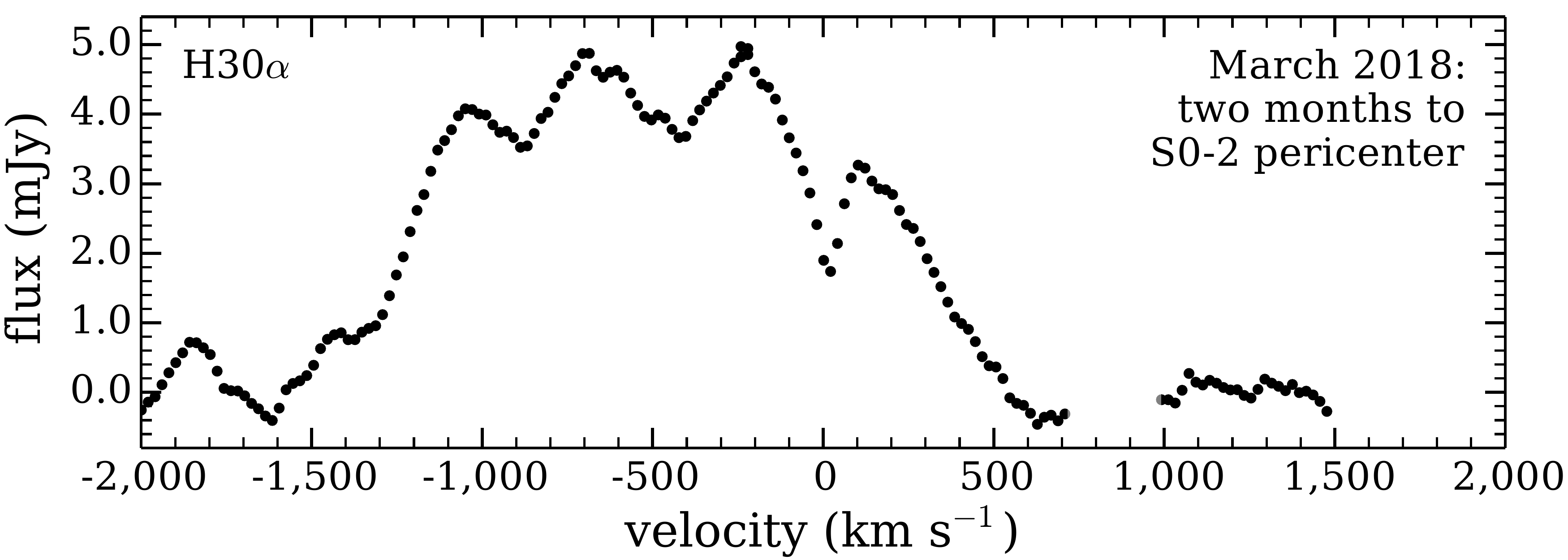} \\
\includegraphics[width=0.7\columnwidth]{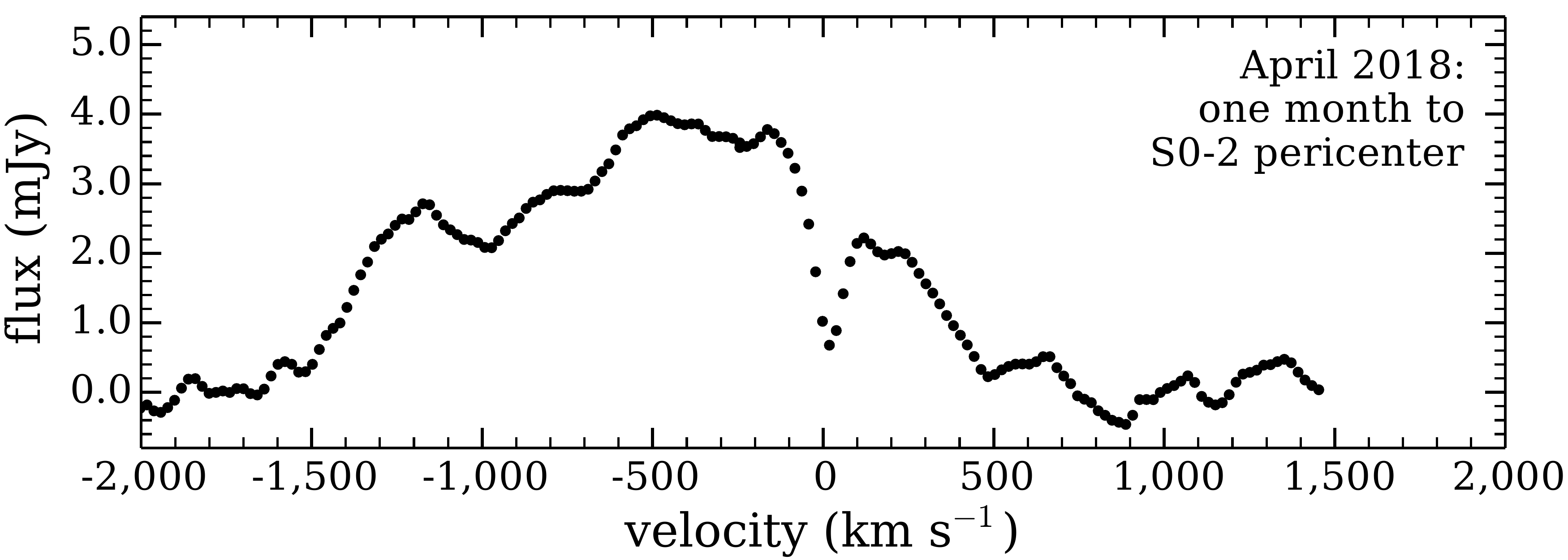} \\
\includegraphics[width=0.7\columnwidth]{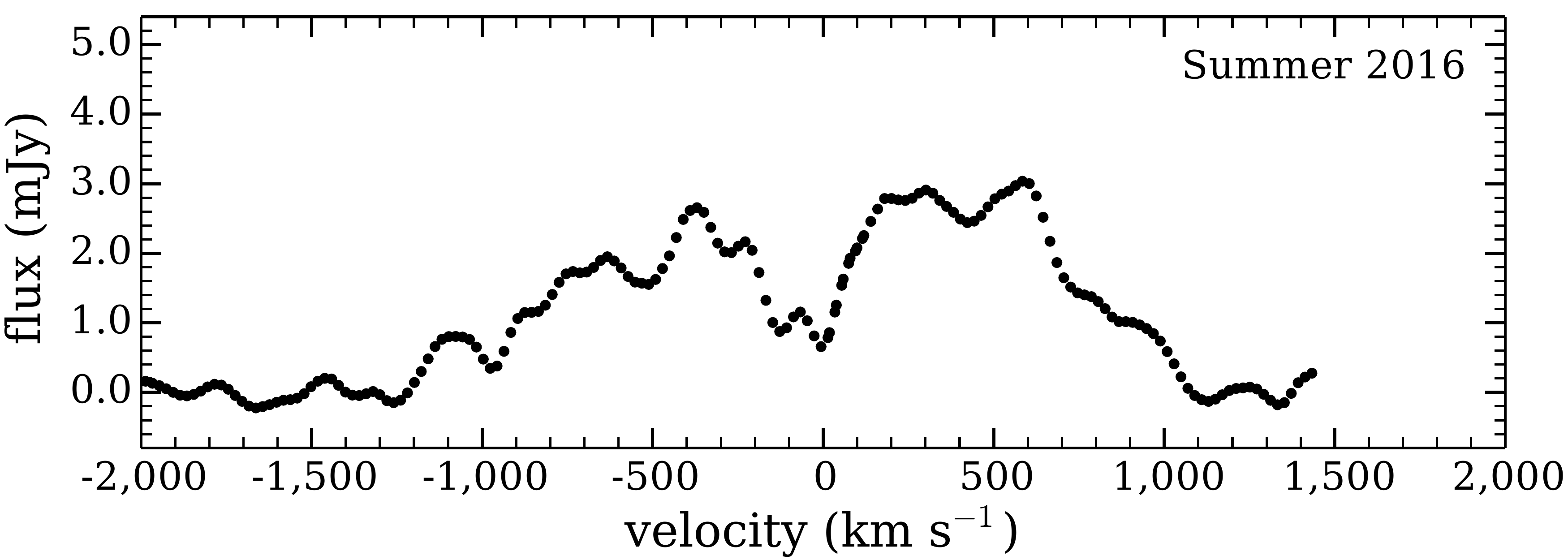} 
\end{tabular}
\caption {
Evolution of the broad recombination line of hydrogen H30$\a$ ($\mathrm{n}=31 \to 30$) obtained from within $1.2\times0.8 \, \arcsec^2$ centered around Sgr A* during the close passage of the S0-2 star. The observations are conducted with ALMA in mid-March and mid-April 2018 and correspond to two and one month prior to the S0-2 star's pericenter passage, respectively. The source of the emission is the cool disk (M19). Continuum emission is subtracted. The 2018 spectrum is boxcar-smoothed by $150 \, \kms.$  Contamination by extended emission from outside the region of interest is removed (Appendix \ref{app:dip}). The details of the observations and the uncertainties are presented in Section \ref{sec:analysis}. For comparison, we show the spectrum observed in 2016 and reported in M19. The 2016 spectrum is boxcar-smoothed by $45 \, \kms.$ 
}
\label{fig:spec}
\end{figure}

Figure \ref{fig:spec} shows variation in the spectrum of the broad hydrogen recombination line H30$\a$ from a $1.2\times0.8 \, \arcsec^2$ region centered on Sgr A* during the passage of the S0-2 star. The source of this emission is the cool disk (M19) surrounding Sgr A* at about  $2\times 10^4 R_\sch,$ which is composed of $10^4$~K gas clumps with neutral gas component. The size extraction region is dictated by the largest size of the telescope beam during the observations, it is about twice as large as the size of the cool disk (0.45 arcsec). The extraction region is about the size of the telescope beam during the April observations (1.0 arcsec) and about twice the size of the telescope beam during the March observations (0.5 arcsec).

The total width of the H30$\a$ line emission is about $2000 \, \kms.$ The spectrum extends much further to the blue side than to the red side. The velocity-integrated line fluxes $S\Delta V$ (the integrals under the spectral line) are
\be\label{eq:SdV2018a}
    && [S\Delta V]_\mathrm{2018.03} = 6.1 \pm 1.0 \, \Jy \, \kms \\
    && [S\Delta V]_\mathrm{2018.04} = 4.9 \pm 1.0 \, \Jy \, \kms. \label{eq:SdV2018b}
\ee

The relative difference between the velocity integrated line fluxes of H30$\a$ between the March and April 2018 observations is
\be\label{eq:dSdV}
    \Delta [S\Delta V]_\mathrm{2018} = 1.2 \pm 0.6 \, \Jy \, \kms.
\ee
In calculating the uncertainty of the value of $\Delta [S\Delta V]_\mathrm{2018}$ we used the following. The contributions of the potential foreground molecular line emission to the March and April spectra must be nearly identical and cancel out during the subtraction. The systematic errors due to spectral window alignment affect both spectra in the same way, and therefore largely cancel out during the subtraction. 
We consider eq. (\ref{eq:dSdV}) a conservative estimate of the change of the flux between the two epochs because of the possible leakage of emission from large radii (Appendix \ref{app:dip}). The flux variation between the two epochs could be larger by up to $0.2 \, \Jy \, \kms,$ caused by an increased contribution of external leakage emission in April due to larger beam size. It is not possible to take this into account with the available data.

\section{Discussion} \label{sec:discussion}

\begin{figure*}
    \centering
    \begin{tabular}{ccc}
    \includegraphics[width=0.3\textwidth]{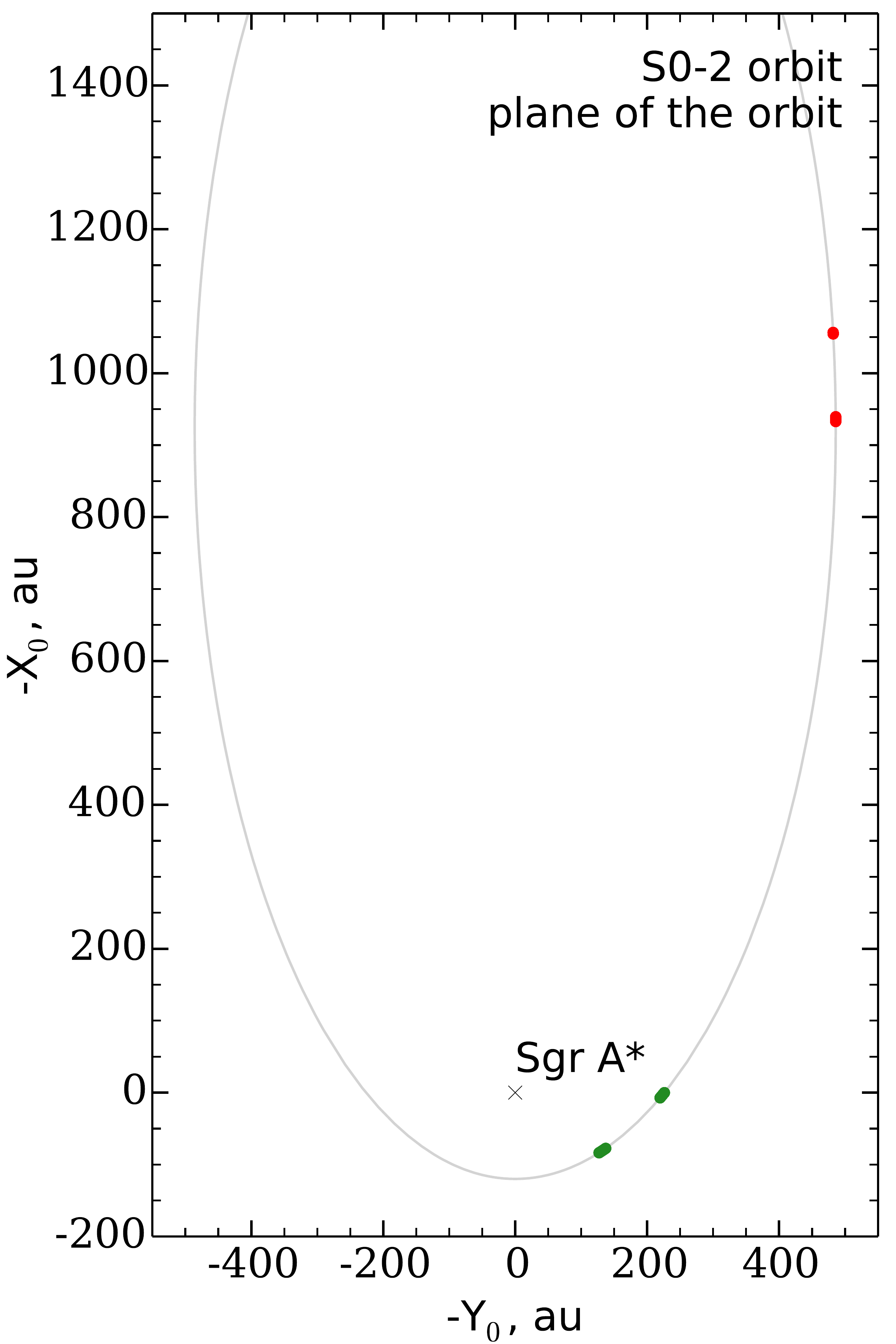} &
    \includegraphics[width=0.3\textwidth]{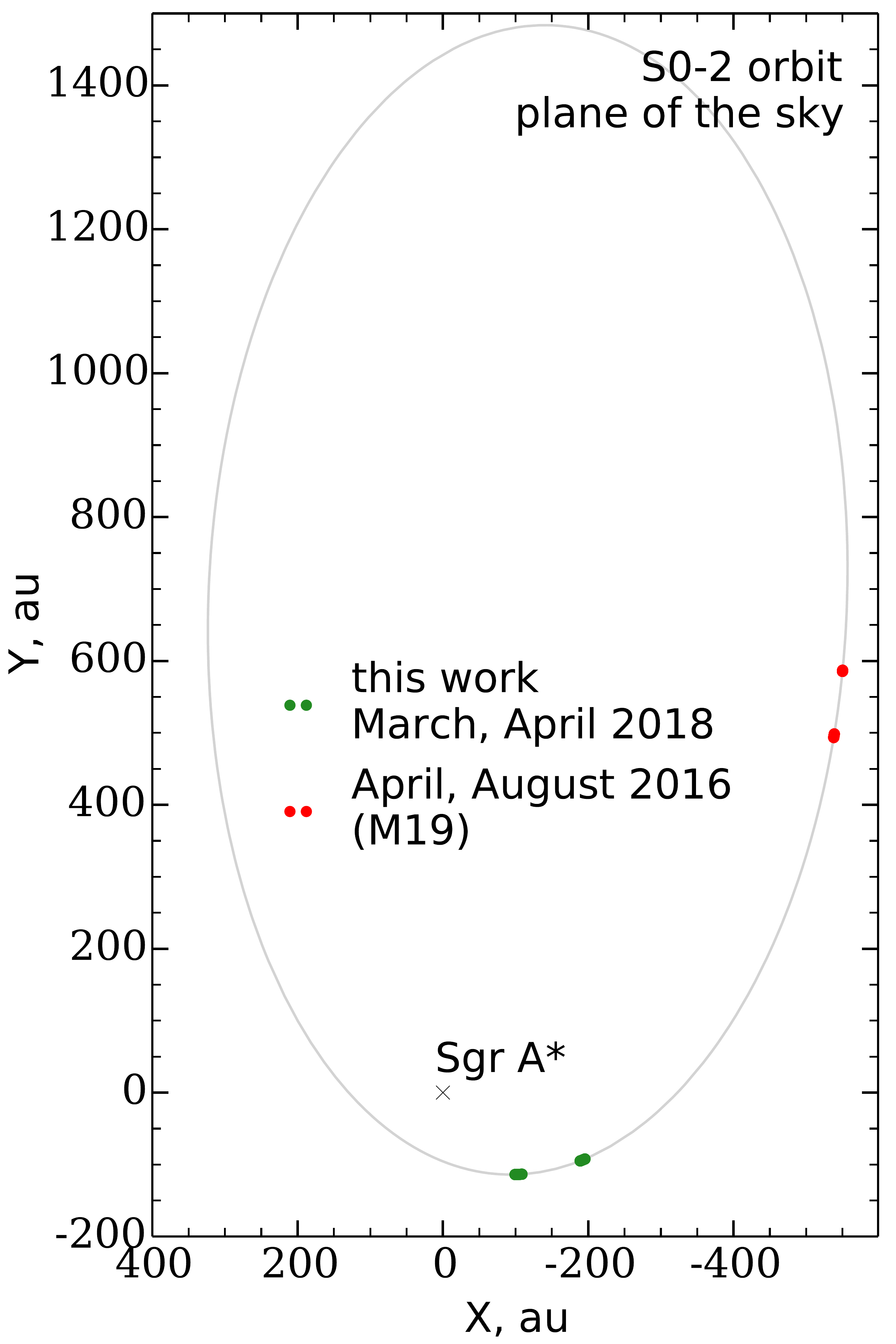} &
    \includegraphics[width=0.3\textwidth]{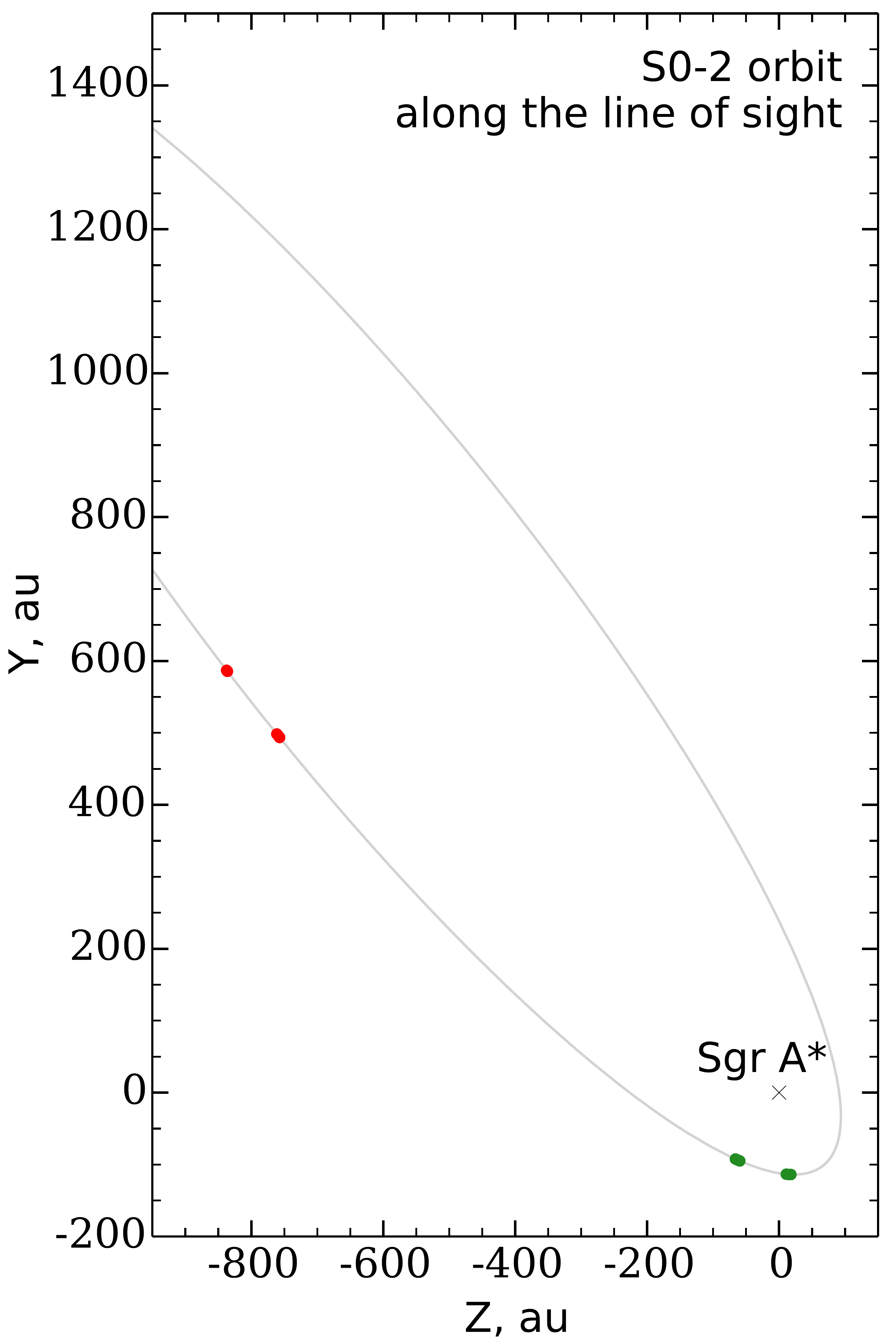}
    \\
    \end{tabular}
    \caption{Trajectory of the S0-2 star around the Galactic Center black hole with its location during the current 2018 observations (green) and M19's observations (red). The black cross marks the location of the Sgr A$^*$ at $(0,0)$. Coordinates $(X,Y)$ correspond to relative (RA, Dec) offsets. The $Z$ axis points away from the observer. The negative $Z$ are between us and Sgr A*, and the positive $Z$ are behind the black hole. Left panel: The orientation in the plane of its orbit. Coordinates are rotated to $(-Y_0, -X_0)$ for visualization purposes. Middle panel: Projection on the plane of the sky. Right panel: The projection on the plane perpendicular to the plane of the sky. 
    The details of coordinate system transformations are in Appendix \ref{app:coord}.
    At the distance of the Galactic Center -- 8 kpc -- 1 arcsec corresponds to about $8000$ au. 
    }
    \label{fig:traj}
\end{figure*}

In this Section we compare the results with M19 and discuss an interpretation of the recombination line variation.
The total widths of the spectra of hydrogen recombination H30$\alpha$ line emission in Figure \ref{fig:spec} is $2000 \, \kms$ which is the same as in the 2016 observations reported in M19. The blue-shifted wing of the spectrum in both March and April 2018 datasets is wider than the red-shifted wing, while the M19's spectrum had two mostly equal red-shifted and blue-shifted wings.

As seen in the observer's image plane, the S0-2 star approached Sgr A* on the blue-shifted side of the cool disk (with respect to the observer). We attribute the increase in the strength of the emission on the blue-shifted side to the ionization of pockets of neutral gas within the cool disk cloudlets, with additional ionising photons brought by the passing star. We attribute the changes of the spectral shape to the disk's long term evolution and the disk's interaction with winds of the approaching S0-2 star. 

The velocity-integrated H30$\a$ line flux in 2018 data ($[S\Delta V]_\mathrm{2018.03}$ and $[S\Delta V]_\mathrm{2018.04}$) presented in equations (\ref{eq:SdV2018a}-\ref{eq:SdV2018b}) is larger than the $[S\Delta V]_\mathrm{2016} = 3.8 \pm 0.8 \, \Jy \, \kms$ reported in M19. The absolute values of $[S\Delta V]_\mathrm{2016}$ and $[S\Delta V]_\mathrm{2018.03}$ can be consistent with each other due to large uncertainties. Calculation of the relative difference between the 2016 and 2018 data does not benefit from cancelling out systematics (as in equation \ref{eq:dSdV}), as they are processed using two independent methods.
The long term evolution during the 2016 observations, 2017 observations (not yet published), and 2018 observations will be discussed in Murchikova et al 2022 (in preparation).
Here we concentrate on the relative difference between the March and the April 2018 observations, i.e. $\Delta [S\Delta V]_\mathrm{2018}$ in equation~\ref{eq:dSdV}.

We consider the supply of the ionizing photons to the cool disk and the S0-2 star's contribution to it. The dominant source of ionizing radiation near Sgr A* is the group of  Wolf–Rayet stars (WR) orbiting at $\sim 4$ arcsec \citep{Martins2008}. Most of these stars are in the counterclockwise stellar disk \citep{Levin2003}. The bolometric luminosity of WR stars is $\sim 10^5 - 10^6 L_{\odot},$ of which 39 -- 69\% is emitted as ionizing radiation \citep{Crowther2007}.
Thus the production rate of ionizing photons from an average WR star is  
\be
    Q_0^{WR} \sim \frac{0.55 \times 10^{5.5} L_{\odot} }{13.6 \mathrm{eV}} = 3 \times 10^{49} \, \sec^{-1}.
\ee
Fifteen such stars orbiting at about 4 arcsec produce the flux of $3.8 \times 10^{47}$ ionizing photons per second through the sphere of the radius $R_{disk}=0.23$ arcsec (0.009 pc) centered on Sgr A* and containing the cool disk. To simplify calculations we neglect the existence of the hole in the middle of the disk, as accounting for it leads to a correction which is much smaller than other uncertainties. 

\begin{figure*}
    \centering
    \includegraphics[width=0.75\textwidth]{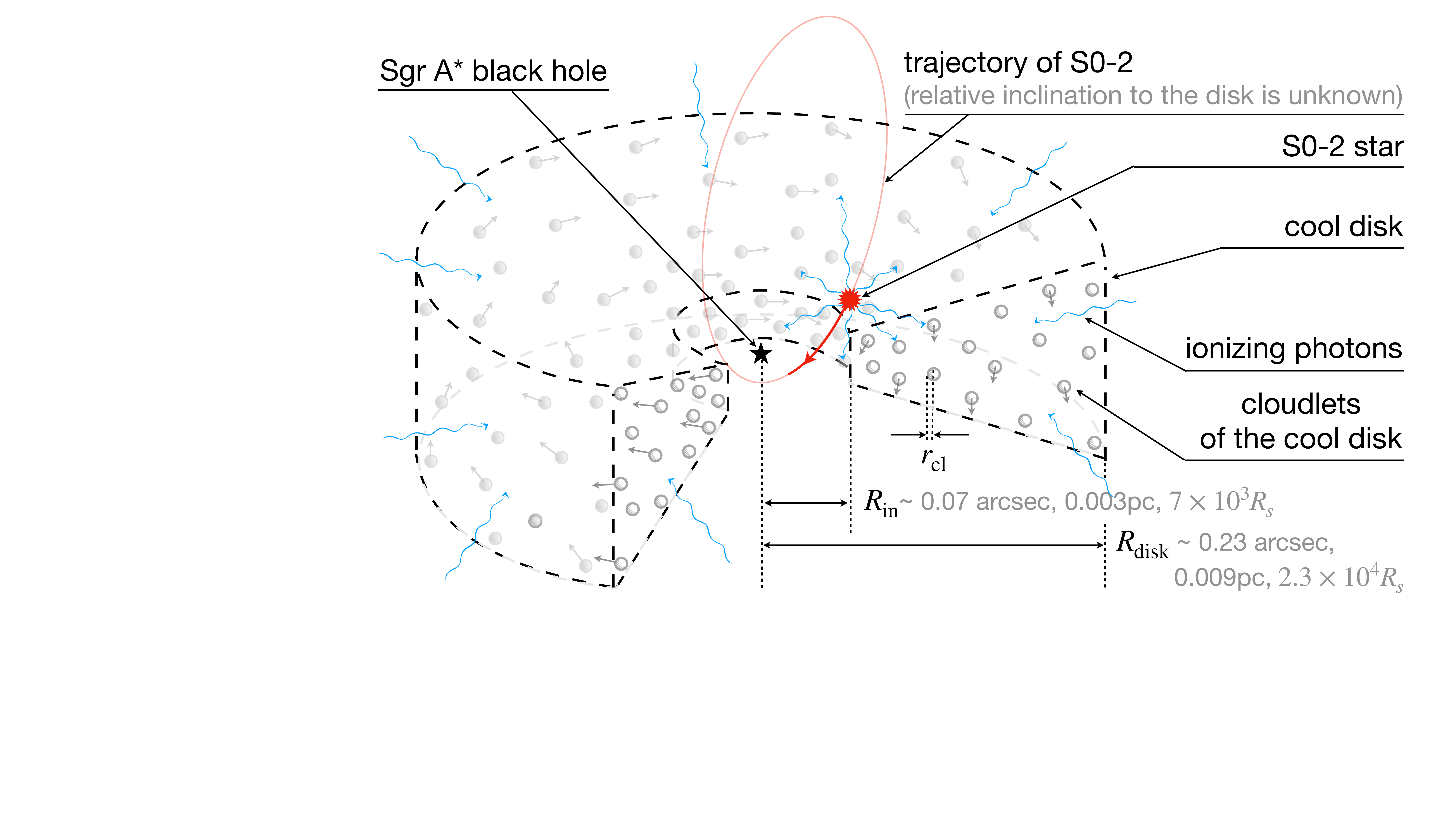}
    \caption{The passage of the S0-2 star by Sgr A*'s cool disk. The black star marks the location of the black hole. The dashed line outlines the cutaway of the cool disk. 
    The red star marks the S0-2 star. The pale red oval is the trajectory of the S0-2. Its relative orientation with respect to the cool disk is unknown.
    Grey circles depict the cloudlets of cool gas which are filling the cool disk. Grey arrows mark the velocities of the cloudlets. For visualization purposes, we omitted the arrows at some of the cloudlets. Wavy blue arrows represent the ionizing photons reaching the disk. They are predominantly supplied by the WR stars at 4 arcsec distance from the Galactic Center, which are outside of the plotted region. The contribution of the S0-2 to the ionizing photon flux is illustrated by wavy blue arrows emanating from the S0-2. During the time of our observations the S0-2 is moving with the velocity of about 3700 $\kms.$ The star's motion during February - April 2018 (starting 1.3 months prior to our first observation and ending on the date of our second observation) is marked by a solid red line with an arrow indicating the direction. 
    The characteristic scales of the cool disk in the units of arcseconds, parsecs and Schwarzschild radii of Sgr A* are indicated in the bottom right corner. The outer radius of the disk is $R_{\mathrm{disk}}.$ It is plotted as $0.23$ arcsec, however, its true extent is $0.11 \mathrm{arcsec} < R_\mathrm{disk}< 0.23 \mathrm{arcsec}.$ Most of the disk's emission is contained within $0.23$ arcsec radius and the peak of the cool disk emission is at 0.11 arcsec  \citep{Murchikova2019}.
    The radius of the hole in the middle of the disk is $R_\mathrm{in}.$ It is plotted as $0.07$ arcsec, however we only know that its size is approximately $0.1$ arcsec. The radius of a typical cloudlet is $r_{cl} \sim R_\mathrm{disk}/500.$ The plot is only partially to scale: the extent of the path  of the S0-2 between the February and April should be about 1.5 times shorter and the typical cloudlet's radius should be much smaller than plotted. The opening angle of the cool disk is unknown.
    }
    \label{fig:cd_scheme}
\end{figure*}

We can calculate the filling factor of the disk $(f_{cd})$ using a model composed of spherical cloudlets with hydrogen number density $n$ and radius $r_{cl}.$ It reads 
\be\label{eq:f_cd}
    f_{disk} &\simeq& \frac{N_{cl} A_{cl}}{A_{disk}} \\
    &=& 0.05 \left( \frac{n}{0.7 \times 10^6 \, \cm^{-3}} \right)^{-2} \left( \frac{r_{cl}}{R_{disk}/500} \right)^{-1}.
\ee
Here $A_{disk} \sim \pi R_{disk}^2$ is the total surface area of the disk, we neglected the hole in the middle (Figure~\ref{fig:cd_scheme}), $A_{cl}=\pi r_{cl}^2$ is the cross section of a cloudlet, and $N_{cl} \simeq M_{disk}/(\frac{4}{3} \pi r_{cl}^3 m_p n)$ is the number of cloudlets in the disk. The total mass of such a disk calculated using the results of M19 is
\be
M^{ion}_{disk} \sim 2.1 \times 10^{-5} \left( \frac{n}{0.7\times 10^6 \, \cm^{-3}} \right)^{-1} ~ M_\sun.
\ee
We assume that the total mass of the disk is similar to the mass of the ionized component ($M^{ion}_{disk}$). The mass calculated above depends only on the $\Delta [S\Delta V]_\mathrm{2018} $ and the density  $n$ within the cloudlets. In this work, we set the magnification of the H30$\a$ line due to continuum pumping to the value of ${\M}={100},$ following \cite{Murchikova2019} and \cite{Ciurlo2021}.

Taking into account estimation for the cool disk's filling factor (eq \ref{eq:f_cd}) and averaging over geometrical inclinations we find that the cool gas clumps receive 
an ionizing photon flux from WR stars equal to
\be\label{eq:Q0wr} 
    Q^{cd,WR}_{0} &\sim&  f_{disk} \, Q_0^{WR} \left(\frac{2}{\pi}\right)^2 \frac{\pi (0.23 \arcsec)^2}{4 \pi (4 \arcsec)^2}\\
    &=& 7.5 \times 10^{45} \sec^{-1}.
\ee
This number does not account for the potential losses the ionizing photons sustain on the way from the WR stars to the Sgr A*. The value $Q^{cd,WR}_{0} \sim Q_0^{cd,M19}\sim 5 \times 10^{45} \, \sec^{-1}$ is in agreement with the estimated flux of ionizing photon flux necessary to keep the gas ionized. 

We now consider the changes to this flux due to the passage of the S0-2 star. The schematics of the star's flyby by the disk is in Figure \ref{fig:cd_scheme}. The S0-2 star is an early B star of spectral type B0-B2.5V, which emits a flux of ionizing photons equal to $Q_0^{S02} \sim 4.7 \times 10^{46} \, \sec^{-1}$ (Appendix \ref{app:S0-2}). During most parts of its trajectory, when it is not embedded in the cool disk and not passing though its center, about $f_{disk}/{2} = 2.5 \%$ of the star's photons are falling into the disk's cloudlets. The factor of one half is due to the fact that, in such a configuration, about half of the star's photons are emitted in the direction opposite to the disk. The ionized photon supply to the cool disk from the S0-2 is then 
\begin{equation}
    Q^{cd,S02}_{0} \sim  f_{disk} \, \frac{Q^0_{S02}}{2} = 10^{45} \, \sec^{-1}. 
\end{equation}
When the S0-2 passes through the disk, the filling factor can increase several times, reaching $\sim 0.2$ (depending on the cloudlet distribution and disk's geometry). The intercepted ionizing flux contributing to the recombination emission can be even greater if the star is passing particularly close to a cloudlet. 

The excess of the velocity-integrated line flux of $\Delta [S\Delta V]_\mathrm{2018} = 1.2 \, \Jy \, \kms$ in March 2018 can be converted into the mass of the ionized gas using the disk model described above (M19). It implies the presence of an extra 
\be
	\Delta M^{ion}_\mathrm{disk} &\sim& 0.66 \times 10^{-5} \left( \frac{n}{0.7 \times 10^6 \, \cm^{-3}} \right)^{-1}  M_{\odot} \\
	&\sim& M_{disk}^{neutral}    \label{eq:dmass}
\ee
worth of ionized gas that was seen in March 2018 but not in April 2018.
We believe that this gas was converted from the neutral phase into the ionized phase with the help of the supply of the ionized photons from S0-2.
To ionize this amount of neutral gas we need to ionize $7.9 \times 10^{51}$ neutral hydrogen atoms, requiring at least the same number of ionizing photons.

The S0-2 is fast-moving: it covered a distance 
equal to about 10\% of the disk's size during the month between our observations. It carries a sizable supply of ionizing photons capable of achieving the required ionization. When the excess of ionizing photon supply diminishes, due to the departure of the S0-2 to a new and less favorable location for radiation capture, the ionized gas starts to recombine back to the neutral state. The characteristic recombination timescale is
\be 
    \t_{rec}\sim \frac{1}{\a(H^0,T) \, n} = 1.3 \, \mathrm{month} \times \left( \frac{n}{0.7 \times 10^{6} \, \cm^{-3}} \right)^{-1}.
\ee 
Here we used the total recombination coefficient $\a(H^0,T=10^4 \, \K) = 4.18 \times 10^{-13} \, \cm^3 \sec^{-1}.$
This is about the same as the time between observations.

The S0-2 does not have to maintain the ionization of neutral gas or consistently supply an excess of ionizing photons.
During its passage through the disk, the star ionizes the parcels of neutral gas that it passes close to. After the star has departed to a new location and the supply of the ionizing photons has diminished, these parcels' recombination is then visible for about 1.3 months. The strongest March spectrum is also the spikiest, having more identifiable peaks than the April data. This suggests that the increase of the velocity-integrated line flux is indeed due to the ionization of a particularly large group of cloudlets. To quantify the spikiness of the spectra, we calculate the mean square deviation of the spectra plotted on Figure \ref{fig:spec} from its average within 400 $\kms.$ For the March spectrum it is 0.28 mJy, which with 1.5 times larger than the 0.19 mJy obtained for the April spectrum.

The $7.9 \times 10^{51}$ ionizing photons required to ionize $M^{neutral}_{disk}$ in equation (\ref{eq:dmass}) can be supplied by the S0-2 within 2 days. This is 5\% of the number of ionizing photons emitted by the star during the characteristic recombination time $t_{rec}.$ Consequently, during the month prior to our March 2018 observations, the S0-2 passed through locations particularly favorable for ionization of neutral gas parcels. Then, between March and April 2018, the S0-2 left the favorable location(s) and a large fraction of the neutral gas recombined back into the neutral state and again, becoming invisible  in the recombination line.


\section{Conclusions}\label{sec:conclusion}

We present ALMA observations of the broad 1.3 mm recombination line of hydrogen H30$\a: \, \mathrm{n} = 31 \to 30$ from the cool disk (M19) surrounding Sgr A* during the close approach of the S0-2 star (Figure \ref{fig:spec}).
The observations took place two and one month before the star's pericenter passage in March and in April 2018 (Figure \ref{fig:traj}), respectively, and allow us to test for the existence of neutral gas within ionized $10^4$ K cloudlets of the cool disk.

We find that the velocity-integrated H30$\a$ emission in March exceeds the one in April by about 20\% (equations \ref{eq:SdV2018a} and \ref{eq:SdV2018b}). This variation is due to a change in the amount of ionized hydrogen within cool disk cloudlets, caused by the passage of the S0-2 star, which is a strong source of ionizing radiation (Figure \ref{fig:cd_scheme}). At some point during the $1.3$ months (the characteristic recombination time) before the March observations, the S0-2 passed though a region with a particularly large covering factor by the cloudlets. 
(Most likely this region was in a mid-plane of the cool disk.)
After the star moved away, the neutral gas started recombining and the strength of the recombination line emission decreased. This caused the velocity-integrated line flux in April to be lower than in March. 

We estimate the amount of neutral gas within $2\times 10^{4} R_\sch$ as
\be
	M_{disk}^{neutral} &\gtrsim& (6.6 \pm 3.3) \times 10^{-6} \\
	&\times& \left( \frac{n}{0.7 \times 10^6 \, \cm^{-3}} \right)^{-1} M_{\odot} .
\ee
We use the inequality sign as an indicator that we do not know whether the observed  amount of ionized neutral gas could have been larger if our observations were  conducted before March 2018. The uncertainties are quite large and account only for observational uncertainty of the value of the velocity-integrated line flux (equation \ref{eq:dSdV}). 

Assuming that the predominant cause of the cloudlets losing their orbital momentum and accreting into the black hole is their collisions, and correcting for the presence of the neutral gas fraction, we find that cool gas contribution to the accretion onto the black hole at $\sim 10^4 R_{\sch}$ is (M19 in Supplementary Information) 
\be
    \dot{M} \sim 2.3 \times 10^{-7} \left( \frac{n}{0.7\times 10^6 \, \cm^{-3}} \right)^{-3}\\ \times \left( \frac{r_{cl}}{R_{disk}/500} \right)^{-1} ~ M_\sun \yr^{-1}.
\ee
Here under the term ``cool'' we mean non-X-ray emitting gas, which includes ionized $\sim 10^4$ K gas and even cooler neutral gas shielded within the $\sim 10^4$ K cloudlets. This gas contributes to the hot phase accretion at smaller radii and influences the structure of the accretion flow. 

\section*{Acknowledgement}
We are grateful to  Nadia Zakamska and Claus Leitherer for comments and suggestions.
We thank Erica Keller, Devaky Kunneriath, Melissa Hoffman and Sarah Woods for the working with us during our virtual face-to-face visit to the NAASC, for helpful discussions of data analysis techniques, and for double checking for us the quality of calibrations. We thank Juergen Ott, Nick Scoville, and Selma de Mink for their contribution at the early stages of this project. 

LM's membership at the IAS is supported by the Corning Glass Works Foundation and the William D. Loughlin Membership. LM and TW acknowledge partial support of NRAO's Student Observational Support program. LM is grateful to Dr. David and Barbara Groce for their kindness and encouragement. Part of this work was performed when LM was at the Aspen Center for Physics, which is supported by National Science Foundation grant PHY-1607611. The participation of LM at the Aspen Center for Physics was supported by the Simons Foundation.

This paper makes use of the following ALMA data:
  ADS/JAO.ALMA\#2017.1.00995.S. ALMA is a partnership of ESO (representing
  its member states), NSF (USA) and NINS (Japan), together with NRC
  (Canada) and NSC and ASIAA (Taiwan) and KASI (Republic of Korea), in
  cooperation with the Republic of Chile. The Joint ALMA Observatory is
  operated by ESO, AUI/NRAO and NAOJ.

The National Radio Astronomy Observatory is a facility of the National Science Foundation operated under cooperative agreement by Associated Universities, Inc.

\vfill
\bibliography{cooldisk}

\appendix
\section{Ionizing Flux of the S0-2 Star}\label{app:S0-2}

S0-2 star is an early B star of spectral type B0-B2.5V with the following properties \citep{Habibi2017}:
\be 
    \log{L_{S0-2}/L_\sun}=4.35, \quad T_{eff}= 28,500 \, \K\\
    M= 13.6 M_\sun, \quad R=5.53 R_\sun, \quad \log{g}=4.10.
\ee 
To obtain an estimate of the ionizing photon production rate of the S0-2 star, we use an expression given in \cite{Lanz_Hubeny2007}:
\be 
    Q_0^{S02} = 4 \pi R^2 \times 10^{q_0} \simeq 4.7 \times 10^{46} \, \sec^{-1}.
\ee
Here the exponent $q_0 (T=28,500K, \, \log{g}=4.10) \simeq 22.4$ is obtained from Table 4 in \cite{Lanz_Hubeny2007}.

\section{Removing contamination by emission at larger radii}\label{app:dip}


Our observations are conducted with a relatively large beam  -- 0.5 and 1.0 arcsec in diameter. 
Consequently, we use a relatively large  extraction region to obtain the H30$\a$ spectra -- $0.8 \times 1.2$ $\mathrm{arcsec^2}$. The minispiral \citep{Tsuboi2017} and the clouds associated with it approach Sgr A* within about one arcsecond of the black hole, resulting in possible contamination of the recombination line spectrum we obtain.


In order to determine whether contamination of the cool disk spectra by extended emission takes place, we do the following: 
\begin{enumerate}

\item We notice that March 2018 observations were conducted with a beam size about twice as small as the observations in April. Using March data we can probe emission down to about 0.5 arcsec. Here we refer to the size of the aperture by diameter, e.g. the 0.5 arcsec aperture is the region with a radius of 0.25 arcsec centered on the black hole.

\item \label{item:appB_2}We construct all spectra by summing three components. The first component is the spectrum of the strong point-source at the location of Sgr A* obtained from u-v model fitting and subsequent extraction. The second component is the collection of point-sources spread across the whole extraction region which are produced by the CLEAN algorithm and stored in the CASA \verb|model| data file. We refer to this component as \verb|model| data. The third component is the average of the residual points (not included into the \verb|model| data) within the chosen aperture. We refer to this component as \verb|residual| data. The second and third components are extended, i.e. they are coming from the whole aperture used. The first component is compact and centered on the location of Sgr A*.

\item
We compare the compact and extended components within the 0.5 arcsec and 1.0 arcsec regions, to identify the emission coming predominantly outside of the region of interest. The cool disk is fully contained within the 0.5 arcsec aperture, so the emission coming from the outside this region is likely a contamination. 

\end{enumerate}

\begin{figure*}
\vspace{-0.0cm}
\centering
\begin{tabular}{cc}
  \begin{tabular}{c}
    \includegraphics[width=0.65\columnwidth]{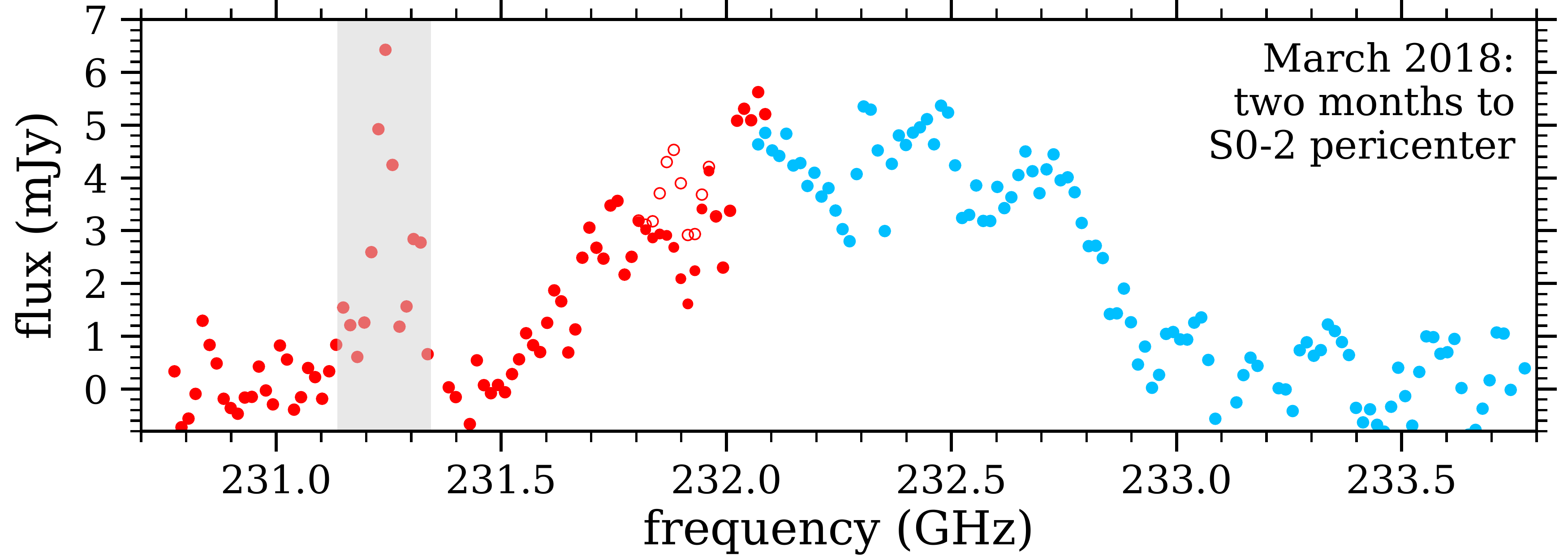}\\
    \includegraphics[width=0.65\columnwidth]{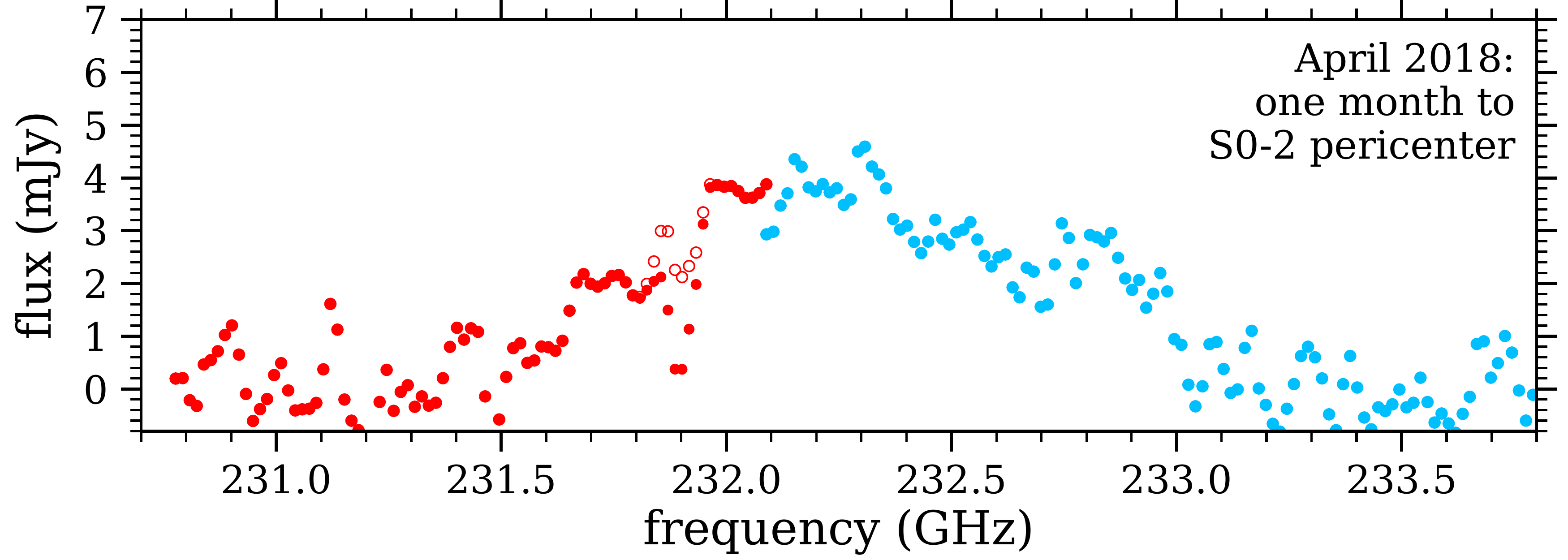}\\
  \end{tabular}
& 
  \begin{tabular}{c}
    \includegraphics[width=0.19\columnwidth]{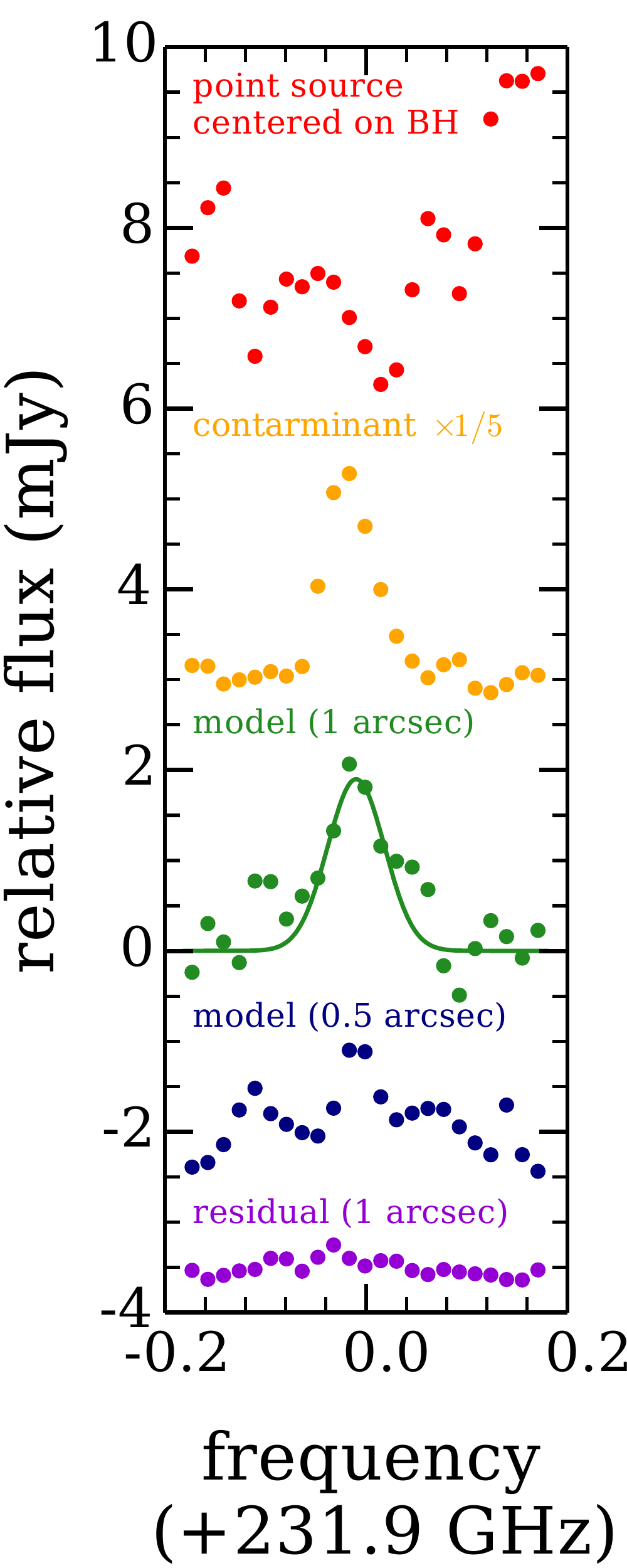}\\
\end{tabular}\\
\end{tabular}
\caption {Left: Spectra of the H30$\a$ emission by the cool disk within the $0.8 \times 1.2$ $\mathrm{arcsec}^2$ aperture centered on the Galactic Center black hole. No spectral averaging is applied. The observations were conducted in March and April 2018, i.e. two and one month before the pricenter passage of the S0-2 star. Different colors represent different spectral windows. The empty circles are spectral point before the minispiral contamination subtraction and filled circles are these after the subtraction. Right: Reasoning for removing contamination by minispiral. The spectrum of the u-v model fitted point-source at the location of Sgr A* (red), and the sum of the CLEAN's model points within the 0.5 arcsec aperture (blue) and the 1.0 arcsec aperture (green) centered on the black hole. Yellow shows the spectrum$\times 1/5$ of a cloud to the South-West of the Sgr A*. The spectrum of the residual is in violet. The separation of the spectra along the Y-axis is random. The frequency range presented is equivalent to $\pm 250 \, \kms$ around H30$\a,$ which is about the width of the minispiral's emission. 
}
\label{fig:by_spw}
\end{figure*}

In the left two panels of Figure \ref{fig:by_spw} we show spectra of the March and April data, color coded by spectral windows and without averaging.
In the right panel of Figure \ref{fig:by_spw} we show the contributions of each component described in point \ref{item:appB_2} to the spectrum of the H30$\a$ line in March 2018. The frequency range shown is equivalent to $\pm 250 \, \kms$ around H30$\a.$  Spectra are vertically offset from each other for display purposes. Red marks the spectrum of the point-source centered on the Galactic Center black hole. Green marks the \verb|model| data within the 1 arcsec aperture. Blue marks the \verb|model| data within the 0.5 arcsec aperture. Violet marks the \verb|residual| data within the 1 arcsec aperture.

The spectrum of the central point-source (red in Figure \ref{fig:by_spw}) has a dip at the frequency of H30$\alpha,$ which is about 2 mJy deep and about 300 $\kms$ wide. A similar dip within the 0.45 arcsec aperture was reported by \cite{Murchikova2019}. The extended \verb|model| emission within the 0.5 arcsec aperture (blue) shows very little excess within the frequency range of the dip. In contrast, the emission within the 0.5 arcsec aperture (green) shows a prominent 2 mJy excess.
This shows that the narrow line excess emission near 230.9 GHz comes predominantly from outside of the 0.5 arcsec aperture (i.e. outside the 0.25 arcsec radius surrounding Sgr A* associated with the cool disk). Consequently, this excess is likely due to contamination. 
The dominant source of the contamination is a cloud to the southwest of Sgr A*. Its spectrum (reduced by a factor of $5$) is plotted in yellow in the right panel of Figure \ref{fig:by_spw}. 
The peak and the width of the contaminant's emission coincide with those of the extended \verb|model| emission within the 1 arcsec aperture (green).

The narrow-line contamination near the H30$\a$ line needs to be corrected for. In order to remove the contamination, we first fit the \verb|model| spectrum within the $1.0$ arcsec aperture to a Gaussian line profile (the solid green line in the right panel of Figure \ref{fig:by_spw}). Then we subtract the fit from the total spectra of the H30$\alpha$ line. The contribution of the \verb|residual| to the combined emission is negligible, so we neglect it in deriving the fit.
We apply the contamination correction to both March and April data. 
The result is presented in the left panel of Figure \ref{fig:by_spw} (filled and empty circles).


In Figure \ref{fig:spec} we removed the points contained in the frequency range shaded in grey in Figure \ref{fig:by_spw}. 
The spurious narrow-line emission at about 231.25 GHz is present in the March 2018 data but not in the April 2018 data. We believe it is caused by one of the calibrators. To produce the spectra we combined observations from multiple executions which use different set of calibrators, and presently we were not able to identify the exact calibrator responsible for the spurious feature. 
The data observed in 2016 and used for the \cite{Murchikova2019} analysis does not have this narrow-line. However, it is present in the data observed in 2017 (currently in preparation).


\section{Rotation transformations from orbital plane into RA and Dec}\label{app:coord}

Let $(X_0, Y_0)$ be orbital plane coordinates of an elliptical trajectory 
\begin{equation}
	\left(
	\ba{c}
		X_0\\
		Y_0
	\ea
	\right)=
	\left(
	\ba{c}
		\size \frac{a(1-e^2)}{1+e \cos \varphi} \cos \varphi\\
		\size \frac{a(1-e^2)}{1+e \cos \varphi} \sin \varphi
	\ea
	\right),
\end{equation}
where $e$ is eccentricity, $a$ is a semi-major axis, and $\varphi$ is the orbital phase angle. We choose parametrization such that pericenter is at $\varphi=0,$ and the star is moving counterclockwise with increasing $\varphi$.

The equatorial coordinate system is a system $(X, Y, Z)$ where the
$X$-axis corresponds to the RA-axis, the $Y$-axis corresponds to the Dec-axis, and the $Z$-axis is perpendicular to the plane of the sky and pointing away from the observer. To rotate $(X_0, Y_0)$ into the equatorial coordinate system we apply the following transformation
\begin{equation}
	\left(
	\ba{c}
		X\\
		Y\\
		Z
	\ea
	\right)=P_3 P_2 P_1
	\left(
	\ba{c}
		X_0\\
		Y_0\\
		0
	\ea
	\right).
\end{equation}
Here the rotational matrices are
\be
P_1=	
	\left(
	\ba{cccc}
		\cos(\w) & \quad -\sin(\w)  & \quad 0\\
		\sin(\w) & \quad \cos(\w)  & \quad 0\\
		0 & \quad  0 & \quad 1
	\ea
	\right), 
\qquad 
	P_2=	
	\left(
	\ba{ccc}
		1 & \quad 0 & \quad 0 \\
		0 & \quad \cos(\pi +i) & \quad -\sin(\pi +i) \\
		0 & \quad \sin(\pi + i) & \quad \cos(\pi+i) 
	\ea
	\right), \\
	P_3=	
	\left(
	\ba{ccc}
		\sin(\W) & \quad -\cos(\W) & \quad 0\\
		\cos(\W) & \quad \sin(\W) & \quad 0\\
		0 & \quad 0 & \quad -1
	\ea
	\right),
\ee
and the rotational angles are parametrized such that $i$ is the inclination, $\W$ is the longitude of the ascending node, and $\w$ is the argument of the pericenter.


\label{lastpage}
\end{document}